\pdfoutput=1
\documentclass{packages/siamart190516}
\usepackage{amssymb}
\usepackage{graphicx,epstopdf}
\usepackage{changepage}

\author{Damyn M. Chipman}
\title{Overview of Solution Methods for Elliptic Partial Differential Equations on Cartesian and Hierarchical Grids}

\begin{document}
\maketitle

% =================================================================================================
% Sections
% =================================================================================================
% \input sections/abstract
% \input sections/intro
% \input sections/swe
% \input sections/elliptic
% \input sections/conclusion

\begin{abstract}

Elliptic partial differential equations (PDEs) arise in many areas of computational sciences such as computational fluid dynamics, biophysics, engineering, geophysics and more. They are difficult to solve due to their global nature and sometimes ill-conditioned operators. We review common discretization methods for elliptic PDEs such as the finite difference, finite volume, finite element, and spectral methods and the linear systems they form. We also provide an overview of classic to modern solution methods for the linear systems formed by these discretization methods. These methods include splitting and Krylov methods, direct methods, and hierarchical methods. Finally, we show applications that would benefit from fast and efficient solvers for elliptic PDEs, including projection methods for the incompressible Navier-Stokes equations and the shallow water wave equations with dispersive corrections.

\end{abstract}

\section{Introduction}

Many physical systems can be described using differential equations. Examples include Newton's Laws of Motion, almost all conservation laws (mass, momentum, energy, etc.), many engineering applications, and many more. To simulate or model such systems, we often use numerical techniques to discretize and solve the corresponding problem on computers. As computational resources are finite, researchers must employ efficient algorithms to solve these problems.

Elliptic partial differential equations are a class of PDEs that explain global and steady state phenomena. Elliptic PDEs arise in fluid dynamics, heat transfer, electromagnetism, geophysics, biology, and other application areas. Many elliptic problems can be solved efficiently, but problems on complex geometries or complicated meshes are more challenging to solve. Solution methods for elliptic PDEs is a well studied topic, with many papers, books, and courses detailing their methods. However, fast and efficient ways to solve elliptic PDEs is an ongoing field of research. Improvements on classical methods, as well as new ideas, have been recently introduced.

\subsection{Outline}

We will first address Laplace's and Poisson's equation in Section \ref{sec:elliptic} as well as numerical approaches to solving these problems. In Section \ref{sec:methods}, we will look at various solution methods for solving the linear system formed by the discretization methods discussed in Section \ref{sec:elliptic}. Next in Section \ref{sec:adaptive}, we look at how elliptic solvers can be used on unstructured and adaptive meshes. In Section \ref{sec:applications}, we give a brief overview of the pressure Poisson equation needed to maintain divergence free velocity fields and a shallow water equation model with dispersive corrections. Finally, in Section \ref{sec:open} we look at open research questions in the area of solvers for elliptic PDEs.

\section{The Elliptic Partial Differential Equation}
\label{sec:elliptic}

Partial differential equations are classified by their highest order derivative terms. A second order, linear differential equation can be written as
\begin{align*}
    A(x,y) \frac{\partial^2 u(x,y)}{\partial x^2} + B(x,y) \frac{\partial^2 u(x,y)}{\partial x \partial y} + C(x,y) \frac{\partial^2 u(x,y)}{\partial y^2} + ... \\
    ... + D(x,y) \frac{\partial u(x,y)}{\partial x} + E(x,y) \frac{\partial u(x,y)}{\partial y} + F(x,y) u(x,y) + G(x,y) &= 0.
\end{align*}
Second-order linear PDEs are classified according to the value of the determinant of this expression:
\begin{align*}
    B^2 - 4AC &< 0,\ \ \ \text{Elliptic} \\
    B^2 - 4AC &= 0,\ \ \ \text{Parabolic} \\
    B^2 - 4AC &> 0,\ \ \ \text{Hyperbolic}
\end{align*}

In this overview, we will consider common elliptic PDEs such as Laplace's Equation
\begin{align}
    \nabla^2 u(x,y) &= 0,
    \label{eq:laplace}
\end{align}
Poisson's Equation
\begin{align}
    \nabla^2 u(x,y) &= f(x,y),
    \label{eq:poisson}
\end{align}
and the variable coefficient Poisson's Equation
\begin{align}
    \nabla \cdot \Big( \beta(x,y) \nabla u(x,y) \Big) &= f(x,y),
    \label{eq:variable_poisson}
\end{align}
where $\nabla = (\frac{\partial}{\partial x}, \frac{\partial}{\partial y})$, $\nabla^2 = \nabla \cdot \nabla = \frac{\partial^2}{\partial x^2} + \frac{\partial^2}{\partial y^2}$, $x,y \in \Omega$, and each with the appropriate boundary conditions on the domain boundary $\partial \Omega = \Gamma$. Such boundary conditions (BCs) can either be Dirichlet (Type-I), Neumann (Type-II), or Robin/Mixed (Type-III) BCs. Dirichlet problems impose the value of $u$ on the boundaries, Neumann problems impose the flux or normal gradient $\partial_n u$ on the boundaries, while Robin problems impose a linear combination of Dirichlet or Neumann type BCs.

\begin{figure}
    \centering
    \includegraphics[width=0.8\columnwidth]{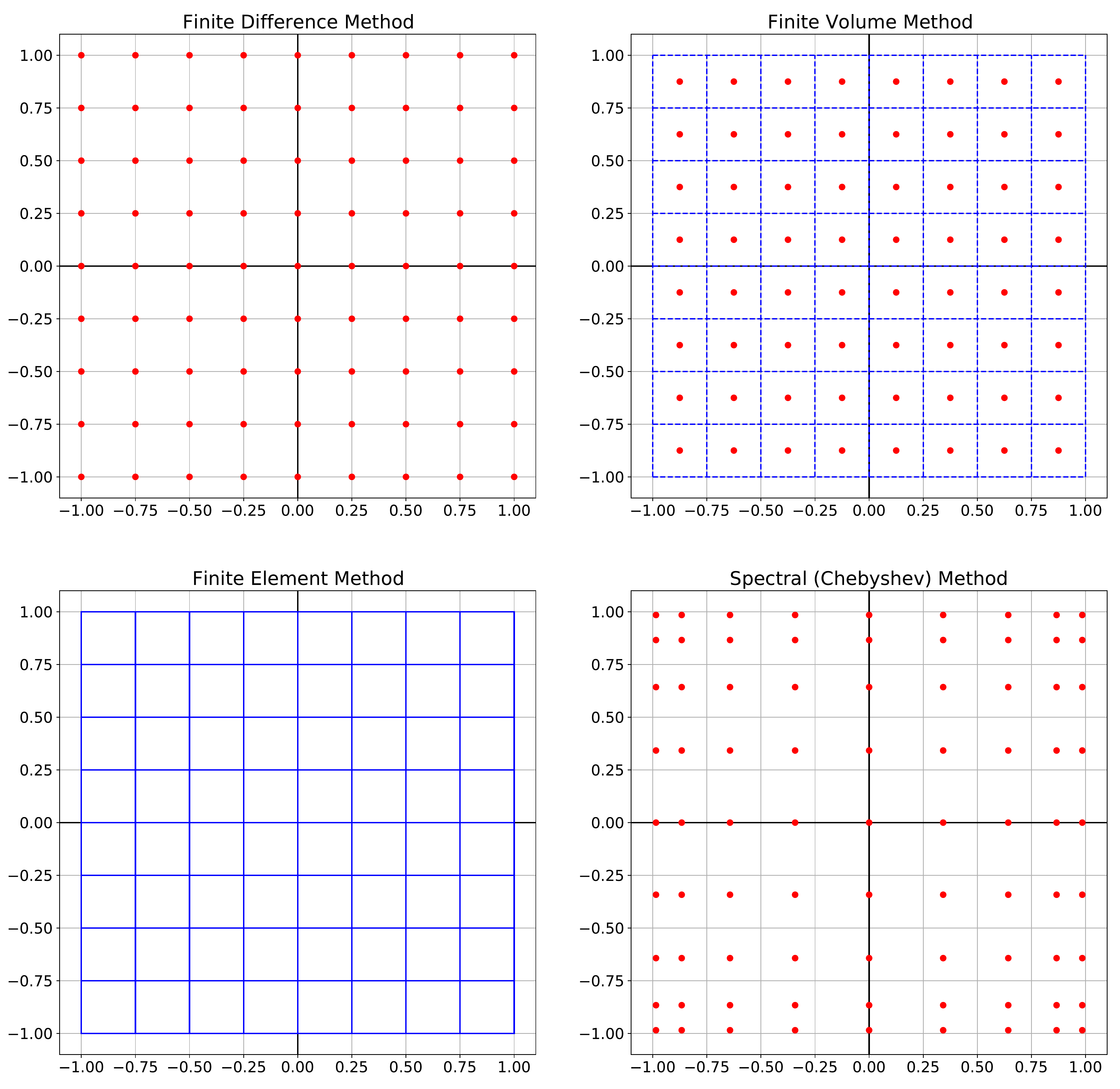}
    \caption{Discretization Methods. Top Left: Finite difference grid based on a mesh of points or nodes. Top Right: Finite volume mesh made of cells (dashed blue boxes) with cell averages in the center (red points). Bottom Left: Finite element mesh where each blue box is an element with a shape function defined at all points within the element. Bottom Right: Spectral mesh made up of a tensor of Chebyshev nodes.}
    \label{fig:discretization_methods}
\end{figure}

Although analytical solutions exist for some (simple) variations of the problems above, we are interested in looking at numerical methods to solving these equations. To use such numerical methods, we first look at various ways to discretize the domain of elliptic PDEs. Common ways to numerically solve elliptic PDEs start with a discrete representation of the domain. We will look at node based, cell based, element based, and spectral based approaches. These approaches are called the finite difference method, finite volume method, finite element method, and the spectral method, respectively. We will refer to Figure \ref{fig:discretization_methods} for a diagram of each.

\subsection{Finite Difference}

If $\Omega$ is on a logically rectangular domain (i.e. 2D Cartesian plane), perhaps the simplest discretization of the domain $\Omega$ is by collocating a mesh of points throughout the domain. Given upper and lower bounds in both directions, $[x_l, x_u], [y_l, y_u]$, the x- and y-location of each point can be defined as
\begin{align}
    x_i &= x_l + i \Delta x,\ \ \ i = 0, 1, ..., N_x-1 \\
    y_i &= y_l + j \Delta y,\ \ \ j = 0, 1, ..., N_y-1
\end{align}
where $N_x, N_y$ is the number of points in the x- and y-direction and grid spacing is defined as
\begin{align}
    \Delta x = \frac{x_u - x_l}{N_x - 1} \ \ \ \Delta y = \frac{y_u - y_l}{N_y - 1}.
\end{align}
These points are shown in the first plot of Figure \ref{fig:discretization_methods}. This leads to the finite difference method, where the function to approximate is solved for at each mesh point. In the finite difference approach, the expressions for the derivatives in a PDE are replaced with Taylor Series approximations. For example, a second order accurate, central-difference approximation to the second derivative can be given as
\begin{align}
    \frac{\partial^2 u}{\partial x^2} &= \frac{u(x - \Delta x, y) - 2u(x, y) + u(x + \Delta x, y)}{\Delta x^2} + \mathcal{O}(\Delta x^2)
\end{align}
and if we let $u(x_i, y_j) = u_{i,j}$, then we can write this as
\begin{align}
    \frac{\partial^2 u}{\partial x^2} &= \frac{u_{i-1,j} - 2u_{i,j} + u_{i+1,j}}{\Delta x^2} + \mathcal{O}(\Delta x^2).
\end{align}
Using these Taylor Series expansions, we can replace the continuous Poisson's equation \ref{eq:poisson} with the discrete system of equations:
\begin{align}
    \frac{u_{i-1,j} - 2u_{i,j} + u_{i+1,j}}{\Delta x^2} + \frac{u_{i,j-1} - 2u_{i,j} + u_{i,j+1}}{\Delta y^2} &= f_{i,j}.
    \label{eq:poisson_FD}
\end{align}
For each grid point $i,j = 0,...,N - 1$, this expression forms a linear system with $(N_x-1) \times (N_y-1)$ unknowns.

\subsection{Finite Volume}

In a finite volume scheme, the domain is broken up into cells, which can be structured or unstructured polygons (2D) or polyhedrons (3D). The function to approximate is solved for in terms of a cell average. Finite volume schemes are often used to solve equations modeling conservation laws, where the cell quantity is conserved in time through balancing fluxes (what comes in and out of the cell boundaries) and the cell source (what the cell is generating or destroying). In a finite volume scheme, the cell average is updated according to the integral formulation of the conservation law as
\begin{align}
    \frac{\partial}{\partial t} \int_{\Omega_i} \textbf{q} d\Omega_i + \int_{\Gamma_i} \textbf{F}(\textbf{q}) \cdot \hat{n}_i d\Gamma_i + \int_{\Omega_i} \textbf{s} d\Omega_i = 0,
\end{align}
where $\Omega_i$ and $\Gamma_i$ are the cell domains and boundaries, respectively, $\textbf{q}$ is a vector of quantities in each cell, \textbf{F} is a flux function, and $\textbf{s}$ is the cell's source term. In the second plot of Figure \ref{fig:discretization_methods}, the blue dashed lines are the cell boundaries, and the red points are the cell centers. For up to second order schemes, the cell average is collocated at the center of the cell; this changes for higher order schemes.

\subsection{Finite Element}

In a finite element scheme, the domain is broken into elements. These elements can also be structured or unstructured polygons (2D) or polyhedrons (3D) as in the finite volume method. The PDE to solve is converted into a weak form by multiplying the PDE by a test function $v(x,y)$ and integrating over the domain. For example, converting Poisson's equation into the weak form looks like this
\begin{align}
    \int_{\Omega} \frac{\partial^2 u}{\partial x^2} v d\Omega + \int_{\Omega} \frac{\partial^2 u}{\partial y^2} v d\Omega &= \int_{\Omega} f v d\Omega \\
    \Rightarrow \int_{\Omega} \frac{\partial u}{\partial x} \frac{\partial v}{\partial x} d\Omega + \int_{\Omega} \frac{\partial u}{\partial y} \frac{\partial v}{\partial y} d\Omega - \Big[ \frac{\partial u}{\partial x} v \Big] \Big|_{x_l}^{x^u} - \Big[ \frac{\partial u}{\partial y} v \Big] \Big|_{y_l}^{y^u} &= \int_{\Omega} f v d\Omega,
\end{align}
where integration by parts is used to transfer a derivative to the test function. The idea behind the finite element method is to use shape or basis functions inside each element (defined over the entire element)
\begin{align}
    u(x,y) \approx \bar{u} = \sum_{k=1}^{N_k} c_k \phi(x,y)
\end{align}
and to use the same basis $\phi$ as the test function $v$ (this is called the Galerkin principle). Upon substitution of the above into the weak form, and integrating the basis function with either quadrature or with analytical expressions, a linear system is formed for the coefficients $c_k$. Often, the coefficients are the actual function values $c_k = u_{i,j}$.

There are several variations of the finite element method. The method explained above forms a global linear system involving all elements. To avoid this, mapping a physical element to some reference element, makes the contributions from a single element only influence itself and its neighbors. This makes the coefficient matrix in the linear system much more sparse and structured, making it easier to solve. Additionally, by not imposing continuity across element boundaries, one arrives at the discontinuous Galerkin method, where the discontinuities are handled through element fluxes. By choosing certain shape or basis functions, one can derive different variations of the finite element method, including the B-spline finite element method \cite{kagan1998new} and the spectral element method \cite{patera1984spectral}.

\subsection{Spectral Methods}

In spectral methods, the approach is to approximate the solution $u$ as a linear combination of a finite set of orthogonal basis functions
\begin{align}
    u(x,y) = \sum_{i=1}^{N_x} \sum_{j=1}^{N_y} c_{i,j} \phi_{i}(x) \phi_{j}(y)
\end{align}
where the coefficients are chosen to minimize a norm, such as the $L^2$ norm of the residual $r(x,y) = \nabla^2 u(x,y) - f(x,y)$. On a discrete domain, this is similar to requiring $\nabla^2 u(x_i, y_j) = f(x_i, y_j)$ at all interior grid points. As this acts like an interpolation scheme, increasing the number of discretization points on with a fixed interval actually leads to highly oscillatory results. Thus, in spectral methods, it is common to use grid points that are clustered near the ends of the interval, such as Chebyshev points, defined on an interval $[a,b]$ as
\begin{align}
    x_i &= a + \frac{1}{2}(b - a)\Big(1 + \cos \big( \pi (1 - \frac{i}{N + 1}) \big) \Big), i = 0, ..., N + 1.
\end{align}
These points are shown on the last plot of Figure \ref{fig:discretization_methods}. With a good basis of grid points, spectral methods can achieve very fast convergence. The linear system formed by this method will be dense as it functions like a high-order interpolation scheme. However, as convergence is much faster than high-order finite difference methods, one can use far fewer points on a grid, so the size of the system is kept small. If one uses Fourier series as the basis functions $\phi$, one can accelerate the solution of the linear system using fast Fourier Transform algorithms. Though, due to the periodic nature of the Fourier series, it is more difficult to implement non-periodic boundary conditions (\cite{leveque2007finite}, \cite{townsend2015automatic}).

\subsection{Other Methods and Summary}

These methods are not all the possible ways to solve an elliptic PDE. Other methods include using quadrature methods on the integral formulation of the PDE, as well as collocation methods such as radial basis function approximations. However, the methods we reviewed show some of the most classic approaches to solving elliptic PDEs.

\section{Solution Methods for Elliptic Partial Differential Equations}
\label{sec:methods}

The discretization methods described above will generally lead to a linear system of equations. The next step is to take advantage of the structure of these linear systems to solve the system efficiently. We will focus on finite difference and finite volume approaches. To generally talk about these solution methods, we assume that we form the linear system
\begin{align}
    \textbf{A} \textbf{u} = \textbf{b}
    \label{eq:ls}
\end{align}
where $\textbf{A} \in \mathbb{R}^{N \times N}$ is a coefficient matrix from one of the discretization methods, $\textbf{u} = u_i = u(x_i)$ for $i \in \textbf{I}_x$ and index set $\textbf{I}_x$ for the discrete domain, and $\textbf{b}$ is a vector with the right hand side information $f$ and encoded boundary information. The goal here is to solve for $\textbf{u}$ where we take advantage of the structure of $\textbf{A}$. We organize the various methods into the following three categories: 1) iterative methods, 2) direct methods, and 3) hierarchical methods.

\subsection{Iterative Methods}

Iterative methods start with an initial guess of a solution to \ref{eq:ls} and correct the iterate until convergence to a specified tolerance. The simplest iterative methods are called splitting methods where the linear system is modified according to
\begin{align}
\textbf{A} = \textbf{M} - \textbf{N} \Rightarrow \textbf{M} \textbf{u} = \textbf{N} \textbf{u} + \textbf{b}.
\end{align}
This suggests the following recursion relationship for the next iteration:
\begin{align}
\textbf{M} \textbf{u}^{k+1} &= \textbf{N} \textbf{u}^k + \textbf{b} \\
\textbf{u}^{k+1} &= \textbf{M}^{-1} \textbf{N} \textbf{u}^k + \textbf{M}^{-1} \textbf{b}.
\end{align}
The idea is to choose $\textbf{M}$ that captures as much of $\textbf{A}$ as possible, but is still easy and quick to invert. As $\textbf{A}$ is either banded or sparse, classical splitting methods for elliptic PDEs split $\textbf{A}$ into $\textbf{A} = \textbf{D} - \textbf{L} - \textbf{U}$, where $\textbf{D}$ is the diagonal components of $\textbf{A}$, and $\textbf{L}$ and $\textbf{U}$ are the lower and upper pieces, respectively. Classical choices for $\textbf{M}$ and $\textbf{N}$ are summarized in Table \ref{tab:splitting}.

\begin{table}[h!]
    \centering
    \begin{tabular}{ | l | l | l |}
        \hline
        Jacobi & $\textbf{M} = \textbf{D}$ & $\textbf{N} = \textbf{L} + \textbf{U}$ \\
        Gauss-Sidel & $\textbf{M} = \textbf{D} - \textbf{L}$ & $\textbf{N} = \textbf{U}$ \\
        Successive Over Relaxation & $\textbf{M} = \frac{1}{\omega}(\textbf{D} - \omega \textbf{L})$ & $\textbf{N} = \frac{1}{\omega} \big( (1 - \omega) \textbf{D} + \omega \textbf{U} \big)$ \\
        \hline
    \end{tabular}
    \caption{Iterative Methods: Splitting Methods}
    \label{tab:splitting}
\end{table}

Another class of iterative methods are called Krylov subspace methods. The Krylov space is defined as
\begin{align}
\mathcal{K}_k = \text{span} \{ \textbf{r}_0, \textbf{A} \textbf{r}_0, \textbf{A}^2 \textbf{r}_0, ..., \textbf{A}^{k-1} \textbf{r}_0 \}
\end{align}
based on the initial residual $\textbf{r}_0 = \textbf{b} - \textbf{A} \textbf{u}^{(0)}$. The goal is to take the next iteration from this particular space. Two common Krylov methods are the Conjugate Gradient method \cite{hestenes1952methods} and the Generalized Minimal Residual (GMRES) method \cite{saad1986gmres}. In the conjugate gradient method, the approximation is adjusted by a conjugate direction, or a vector that is conjugate with respect to $\textbf{A}$. This vector is called the search direction $\textbf{p}$ and is scaled by $\alpha$, which is computed by solving a quadratic minimization problem. The GMRES method builds up an orthogonal matrix $\textbf{Q}$ through a process called the Arnoldi iteration. The Arnoldi iteration forms $\textbf{A} = \textbf{Q} \textbf{H} \textbf{Q}^*$ for orthogonal matrix $\textbf{Q}$ and Hessenberg matrix $\textbf{H}$. The next iteration is found via $\textbf{Q} \textbf{y}$ where $\textbf{y}$ is found from a least squares problem involving $\textbf{H}$. These methods are summarized in Table \ref{tab:ksm}.

\begin{table}[h!]
    \centering
    \begin{tabular}{ | l | l |}
        \hline
        Conjugate Gradient & $\textbf{u}^{(k+1)} = \textbf{u}^{(k)} + \alpha^{(k)} \textbf{p}^{(k)}$ \\
        GMRES & $\textbf{u}^{(k)} = \textbf{Q}^{(k)} \textbf{y}$ \\
        \hline
    \end{tabular}
    \caption{Iterative Methods: Krylov Subspace Methods}
    \label{tab:ksm}
\end{table}

Most iterative methods are considered ``matrix-free" methods. A ``matrix-free" method is a method that does not explicitly form the matrix $\textbf{A}$, but is rather applied to a vector. For example, if implementing a Conjugate Gradient method for a finite difference discretization, one has to compute the product $\textbf{A} \textbf{u}^{(k)}$. Instead of doing the full matrix-vector calculation, one can write a function that takes $\textbf{u}$ and returns the 2nd-order, central difference operator as computed in \ref{eq:poisson_FD}.

As finite difference discretization schemes for elliptic PDEs lead to sparse matrices, the application of $\textbf{A}$ to a vector can be done in $\mathcal{O}(N)$ operations, where $N$ is the number of unknowns in the vector. Thus, the performance for most iterative methods is approximately $\mathcal{O}(N \times N_{iter})$, where $N_{iter}$ is the number of iterations required for a specified tolerance. However, as $N$ gets larger, often so does $N_{iter}$, leading to poor scaling, as noted in \cite{martinsson2019fast}. In addition, iterative methods may not always converge for a given initial guess or structure of $\textbf{A}$, which make them unfavorable for ``black-box" implementations for linear solvers.

\subsection{Direct Methods}

Motivation for direct solvers stems from wanting to improve upon the disadvantages of iterative methods. Martinsson notes in \cite{martinsson2004fast} some advantages to using direct methods over iterative ones:
\begin{itemize}
    \item Direct methods can be applied to multiple right-hand side vectors $\textbf{b}$ or multiple boundary conditions once a factorization or solution operator is built, whereas iterative methods must be solved anew for each right-hand side or for different boundary conditions.

    \item Direct methods can take advantage of ``close" matrices (i.e. if we have an inverse or factorization of $\textbf{A}$ and perturb it by $\epsilon$, we could adjust the inverse to account for it instead of recompute the inverse).

    \item Most direct methods can take advantage of fast and efficient algorithms for matrix factorization such as the singular value decomposition, LU decomposition, QR decomposition, etc.
\end{itemize}
We will look at how direct methods are useful as we consider some common direct methods from \cite{leveque2007finite} and \cite{trefethen1997numerical}.

Many direct methods are based on matrix factorizations. Perhaps the most well-known is the LU decomposition. LU decomposition factors the coefficient matrix into a lower and upper triangular matrix: $\textbf{A} = \textbf{L} \textbf{U}$. The idea is to use Gaussian elimination to eliminate entries below the main diagonal, and then use back-substitution to solve for each entry in $\textbf{u}$. In general, LU decomposition requires $\mathcal{O}(N^3)$ floating point operations and thus is impractical for large matrices. There are banded solvers for Gaussian elimination that can take advantage of the sparsity of a matrix. The Cholesky decomposition is a variant of Gaussian elimination for symmetric matrices. Other matrix factorizations include the QR-decomposition, $\textbf{A} = \textbf{Q} \textbf{R}$, and the singular value decomposition, $\textbf{A} = \textbf{U} \boldsymbol{\Sigma} \textbf{V}^*$.

For a finite difference discretization of elliptic PDEs, $\textbf{A}$ is banded and sparse, and more efficient algorithms for LU decomposition exist. For 1D problems, $\textbf{A}$ is diagonally dominant and sparse with the bandwidth (the number of entries off the main diagonal in a matrix) dependent on the order of the stencil. For the stencil shown in the second order discretization in \ref{eq:poisson_FD}, $\textbf{A}$ is tridiagonal. This allows us to use algorithms such as Thomas's algorithm for solving a tridiagonal system in $\mathcal{O}(N)$ steps (\cite{higham2002accuracy}). In higher dimensions, block versions of Thomas's algorithm exist (\cite{quarteroni2010numerical}).

Compared to iterative methods, direct methods will terminate in a finite number of steps. Direct methods are more fit for ``black-box" implementations. However, because most direct methods need to explicitly form $\textbf{A}$, they are more difficult to implement with limited computing resources. Indeed, going to higher dimensions and higher orders often dramatically increases memory and compute requirements.

\subsection{Hierarchical Methods}

The methods grouped under hierarchical methods attempt to accelerate some of the ideas from iterative and direct methods by breaking the problem into a hierarchy of subproblems. By recursively breaking the original problem into smaller subproblems, significant improvements can be made in complexity and performance. We'll talk about three hierarchical methods here: the multigrid method, nested dissection, and the Hierarchical Poincaré-Steklov (HPS) method.

\subsubsection{The Multigrid Method}

The multigrid method was introduced by Brandt in \cite{brandt1977multi}. It has been widely used in various applications and for all types of solution methods. Briggs in \cite{briggs2000multigrid} gives an overview and tutorial of the multigrid method.

In the multigrid method, the idea is to use multiple levels of grids and solution methods on each level to solve a larger problem. To look at the multigrid method, we define the error in the linear system to be $\textbf{e} = \textbf{u}^{(k)} - \textbf{u}_{exact}$ (the difference between the exact solution and the $kth$ iteration). After a few iterations of a smoother (typically Jacobi's method), because of the local averaging, any high-frequency error is quickly dampened away. What takes longer to eliminate is the low-frequency error associated with the global problem. By coarsening the grid, the lower frequency error is dampened quicker. Multigrid combines the ability of iterative methods to locally reduce error and a coarsening grid technique to accelerate convergence.

\begin{figure}
    \centering
    \includegraphics[width=0.8\columnwidth]{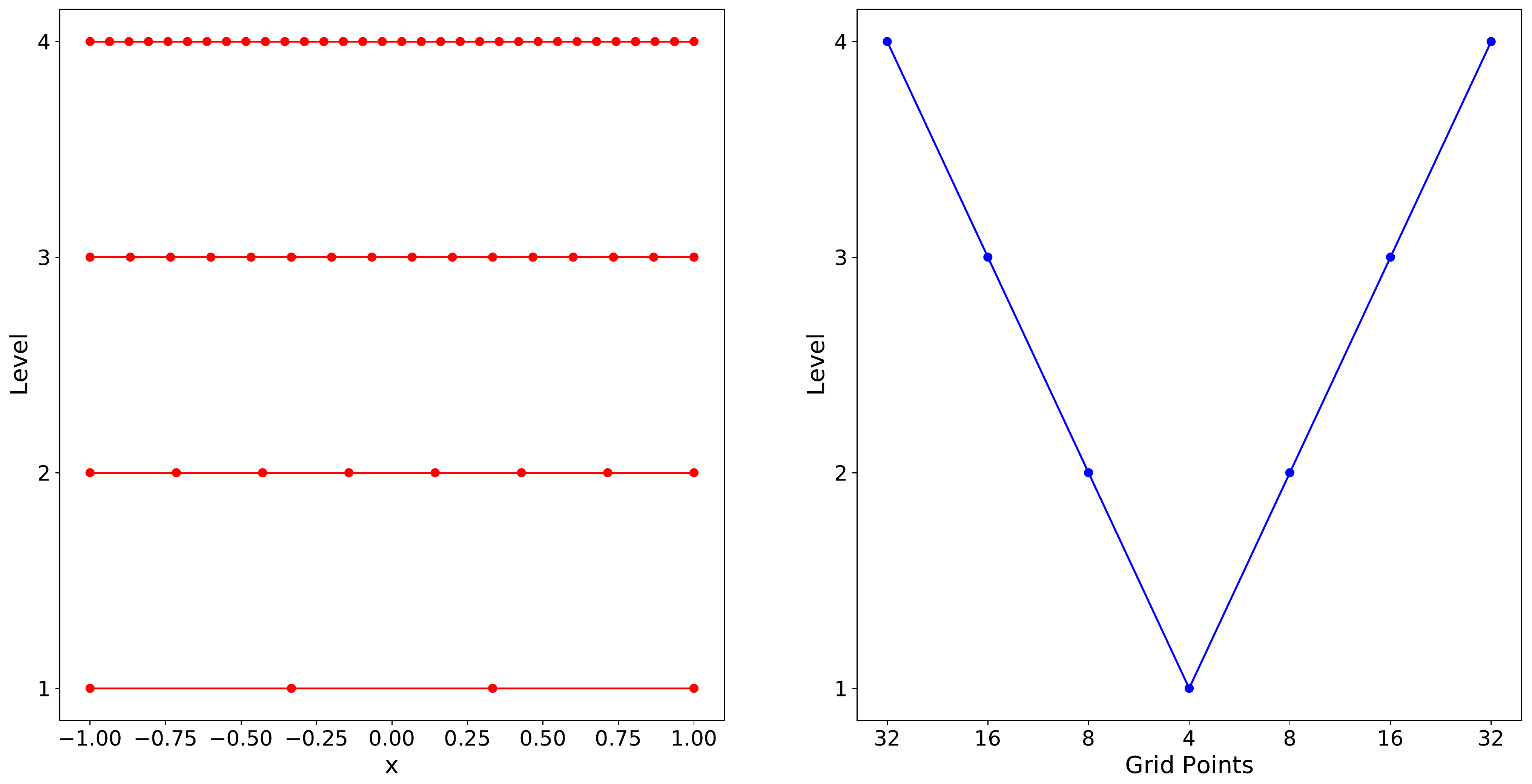}
    \caption{The 1D Multigrid Method. Starting with the finest grid, perform a few iterations of an iterative method. This is called relaxing the solution. Then project onto a coarser grid, and relax again. Do this down the levels in the grid to a desired precision. Once the solution is converged to on the coarsest level, interpolate back up the levels to obtain the solution on the finest level.}
    \label{fig:multigrid}
\end{figure}

In the multigrid method, one starts with the solution on a fine grid, for example, with grid spacing $h$, and performs a few iterations of a smoother. After a few iterations, the $h$-level grid is projected onto a coarser $2h$-level grid. On this level, one performs a few more iterations of a smoother. Because the grid is coarser, this step is faster. Again, after a few iterations, the solution is projected onto an even coarser $4h$-level grid and a few more iterations are performed. This is done a specified number of times, and then the solution is interpolated back up the levels to the original grid. This is shown in Figure \ref{fig:multigrid}.

In what is called the ``full multigrid" method, the process starts on the coarsest level instead of the finest. The solution is relaxed on this level, and then interpolated down onto a finer mesh. The interpolated solution is used as an initial guess for solving the problem on the finer mesh. The relaxed solution on a coarser grid is often an ideal initial guess for the problem on the finer mesh, resulting in quick convergence on that level.

The multigrid method allows one to accelerate an iterative solver. Multigrid methods are very effective as pre-conditioners for iterative methods. This is the general pattern of hierarchical methods: the ability to use classical iterative and direct methods on smaller grids where they perform well, and then ``scale" them up to larger problem sizes.

\subsubsection{Nested Dissection}

The nested dissection method formulated by George in \cite{george1973nested} is a direct method that builds upon Gaussian elimination for problems on a grid. It is also the basis for forming what is called the multifrontal method. By taking advantage of the ordering of points on a grid, one can permute $\textbf{A}$ to first eliminate points that split the mesh into two unconnected meshes. This permutation takes the form of $\textbf{P}^* \textbf{A} \textbf{P}$, where the goal is to form $\textbf{P}$ to reorganize $\textbf{A}$ in a way that eliminates points in an efficient manner.

To detail nested dissection better, consider an $N \times N$ mesh such as a finite difference mesh discussed in Section \ref{sec:elliptic}. Assume that the initial ordering of points corresponds to an index set that iterates over the points in the mesh row-by-row. Now, the idea of nested dissection is to reorganize the points such that we first eliminate points down the middle of the mesh (i.e. the points at $x = 0$ in the first plot of Figure \ref{fig:discretization_methods}). Once these points are solved for, it spilts the mesh into two equally sized pieces that are disconnected. Each of the disconnected meshes now are smaller, and thus easier to solve. This idea of splitting the mesh into disconnected pieces by first eliminating points along an interface can be recursively applied to each split. This means that after dividing the mesh into two, one can divide those two meshes into four meshes, and so on. Recursive splitting is a common characteristic in hierarchical methods.

Nested dissection was first introduced by George in \cite{george1973nested}, and further generalized by Lipton et al. in \cite{lipton1979generalized}. Martinsson has a tutorial on nested dissection in \cite{martinsson2019fast}. In fact, nested dissection served as motivation for another hierarchical method proposed by Martinsson and Gillman called the Hierarchical Poincaré-Steklov method.

\subsubsection{The Hierarchical Poincaré-Steklov Method}

The work done by Gillman and Martinsson in \cite{martinsson2004fast}, \cite{MARTINSSON2013460}, and \cite{gillman2014direct} (with a practical tutorial found in \cite{martinsson2015hierarchical}) culminate in what they call the Hierarchical Poincaré-Steklov (HPS) method. It is a direct solver for elliptic PDEs that is based on a binary tree of rectangular patches where the solution operator to $\textbf{A}$ is built by recursively merging child patches. Like direct methods, the HPS method involves a factorization step and a solve step. The chief advantage of the HPS method over other direct methods is that it does not require the explicit formulation and storage of $\textbf{A}$.

\begin{figure}
    \centering
    \includegraphics[width=0.7\columnwidth]{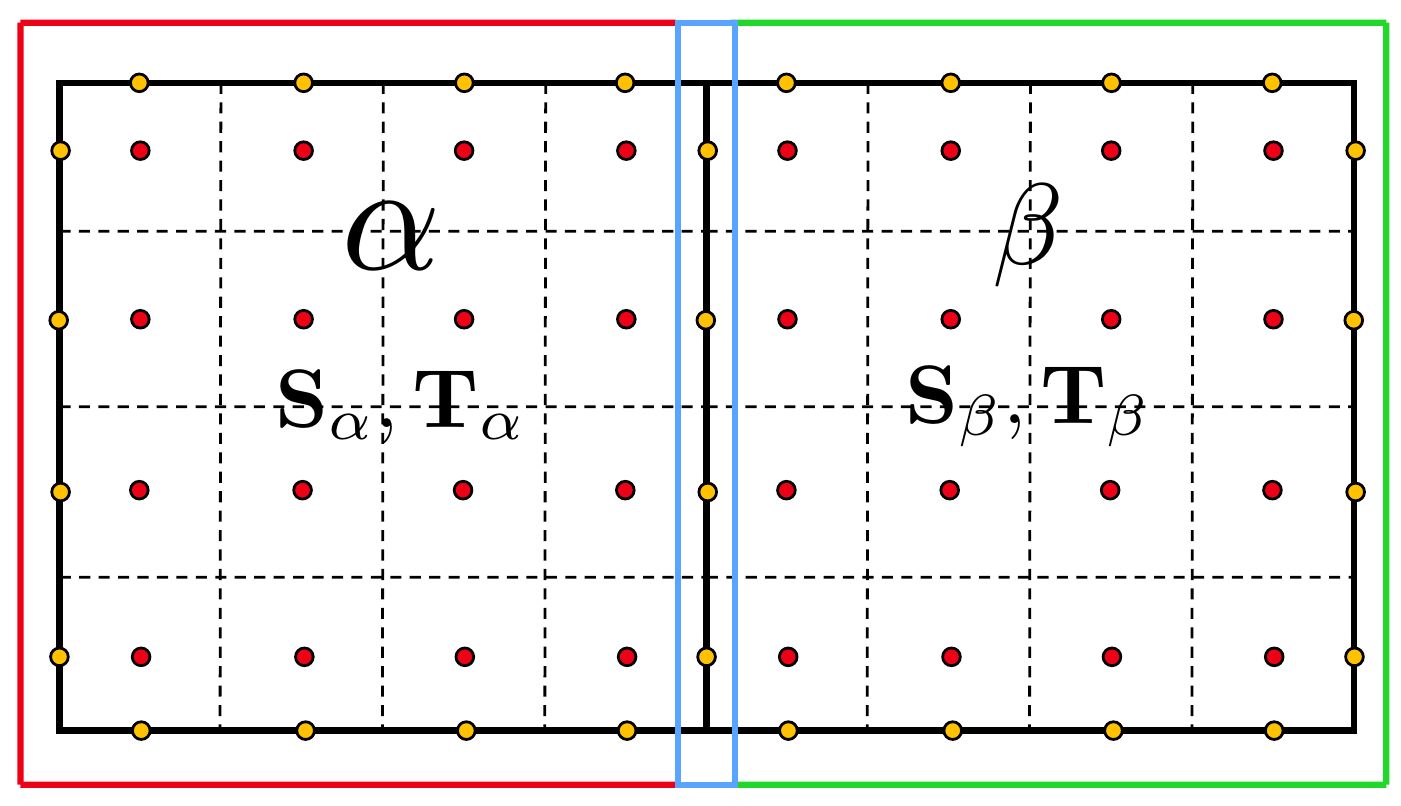}
    \caption{HPS Merge Operation. The merged patch $\Omega_{\tau}$ is the union of children $\Omega_{\alpha}$ and $\Omega_{\beta}$, i.e. $\Omega_{\tau} = \Omega_{\alpha} \cup \Omega_{\beta}$. Red, green, and blue nodes correspond to index sets $\textbf{I}_1$, $\textbf{I}_2$, and $\textbf{I}_3$, respectively. The merge operation eliminates the nodes on the interface of the children patches.}
    \label{fig:merge}
\end{figure}

\begin{figure}
    \centering
    \includegraphics[width=\columnwidth]{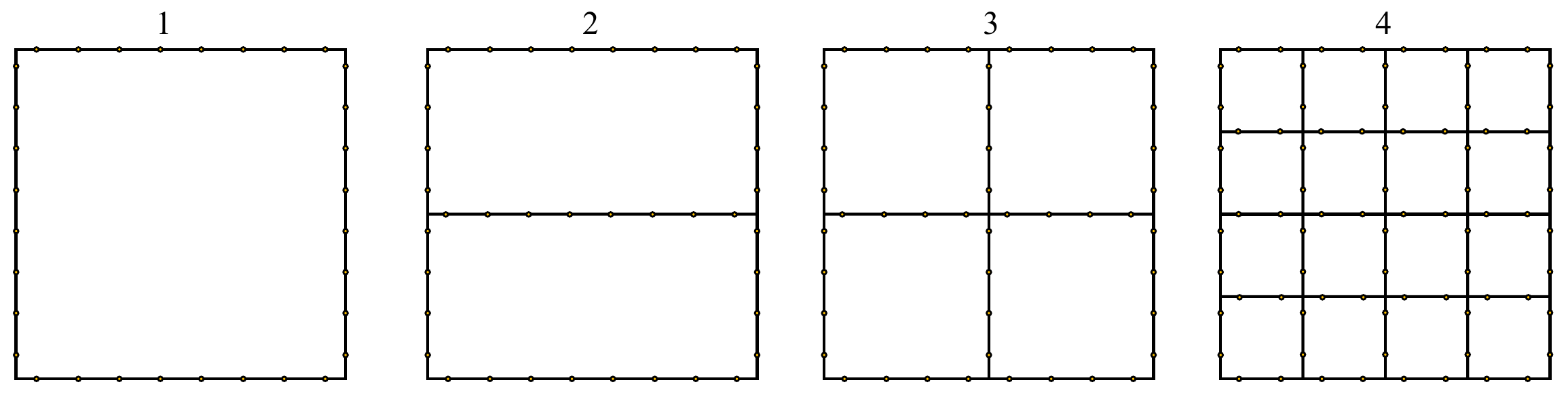}
    \caption{HPS Solve Stage. Once $\textbf{S}_0$ is formed, apply it to the top level Dirichlet data to get boundary (and solution) data on the interface of the children. Apply the patch solution operator down the tree until each leaf has it's local boundary information. Then apply the solution operator to get the solution data in the interior of each leaf.}
    \label{fig:solve}
\end{figure}

The HPS method starts with an original problem domain, and recursively divides the domain in half. This creates a binary tree of patches as shown in Figure \ref{fig:solve}. Once the domain has been decomposed into this tree of patches, two operators are defined on the lowest level, called the leaf level. These operators are the solution operator $\textbf{S}$ and Dirichlet-to-Neumann (DtN) operator $\textbf{T}$. The solution operator maps boundary data to solution data on the interior of the patch (i.e. solves the local boundary value problem), and the DtN operator maps Dirichlet data on the boundary to Neumann data on the boundary. These operators can be formed using any elliptic PDE solver, including fast solvers like spectral methods. After forming these operators, the next step is to recursively merge each sibling patch up the tree. the merge step is demonstrated in Figure \ref{fig:merge}. This results in a global solution operator that can be stored and used multiple times (at different time steps or with varying boundary conditions, etc.), and is similar to a direct method matrix factorization. The final step is applying the solution operator to each level down the tree to obtain the solution everywhere in the domain. This step is just a matrix-vector multiplication and is very fast.

Similar to other direct methods, the HPS method forms an in-memory solution operator that can be applied to several right-hand side vectors. This property makes it ideal for problems where several elliptic solves are necessary. While most iterative methods have better asymptotic performance than other direct methods, the HPS method can be accelerated using hierarchically block seperable (HBS) matrix algebra to achieve near linear asymptotic performance (\cite{gillman2014direct}).

\section{Adaptive and Unstructured Mesh Methods for Elliptic Partial Differential Equations}
\label{sec:adaptive}

In many applications of elliptic solvers where the dynamics of the problem are localized (i.e. shock waves and fronts) or the geometry of the domain is complicated, more complex meshing is used. These types of meshes include dynamically adaptive meshes, where the mesh is refined/coarsened through time, and unstructured meshes, where the cells or elements are typically polygons or polyhedrons and not logically ordered like Cartesian grids. While solving elliptic PDEs on these types of meshes is more difficult, the methods we have talked about can be extended to work on complex meshes.

\subsection{Unstructured Meshes}

When working with unstructured meshes, the domain is broken into polygons (2D) or polyhedrons (3D). Finite difference methods are typically difficult to implement on unstructured grids, but other methods like finite volume and finite element methods are often built around these complex geometries. Just as discretization methods on unstructured grids have been developed, so have more robust solvers. For example, in \cite{luo2006p}, Luo et al. present an unstructured multigrid method for the discontinuous Galerkin method. They use a sequence of solution approximations of different polynomial orders. The ultraspherical spectral element method (ultraSEM) presented by Fortunato et al. is another highly practical extension of the HPS method. In \cite{fortunato2020ultraspherical}, they apply their ultraspherical spectral element method from \cite{olver2013fast} to the HPS method. Fortunato et al. use spectral discretization on mapped quadrilaterals and triangles and use the same merge process from the HPS method. They show that the HPS method also works well on mapped, unstructured meshes.

\subsection{Adaptive Meshes}

Hierarchical meshes, like the meshes used in the HPS method, are also well suited for adaptivity. On adaptive, hierarchical meshes like quadtrees (2D) and octrees (3D), a patch can be recursively split to a desired level of refinement. Areas of the domain that have increased dynamics or need higher refinement (say at boundaries or corners) can be resolved with finer meshes, while other areas can be of lower resolution. This adaptivity increases accuracy without significantly increasing runtime.

In \cite{popinet2015quadtree}, Popinet et al. use a quadtree approach to decompose the domain into a hierarchical grid. In locations where more refinement is necessary, they refine the mesh more. For example, in \cite{popinet2015quadtree}, they are solving the Serre-Green-Naghdi model (we will show this as an application in Section \ref{sec:applications}) which models shallow water equations. As they are looking at a tsunami traveling across the surface of the earth, refinement near the location of the wave fronts is necessary, but without having to have very dense meshing elsewhere. They show that they can implement the multigrid method on a quadtree based mesh.

Additionally, the HPS method has been modified to be used on an adaptive mesh. In \cite{geldermans2019adaptive}, Geldermans et al., they show how to use the HPS method on a quadtree of patches. They demonstrate the merge operation between patches on different levels, as well as discuss proper interpolation schemes for their choice of mesh (a Chebyshev tensor product). In addition, the global solution operator formed by this adaptive HPS method can still be applied to multiple right-hand side problems. The study of fast, direct solvers for elliptic problems on adaptively refined quadtrees and octrees is an active area of research.

\section{Applications of Elliptic Partial Differential Equations}
\label{sec:applications}

We will look at two examples where the use of a fast elliptic PDE solver would greatly benefit the solution process. These are cases where an elliptic solve is required multiple times, such as every time step or for multiple problems.

\subsection{Navier-Stokes: The Projection Method}

The incompressible Navier-Stokes equations is a nonlinear PDE that arises from conservation of momentum and mass in an incompressible fluid. They are expressed as
\begin{align}
    \frac{\partial \textbf{u}}{\partial t} + \textbf{u} \cdot \nabla \textbf{u} &= -\frac{1}{\rho} \nabla P + \nu \nabla^2 \textbf{u}.
\end{align}
In \cite{chorin1967numerical}, Chorin first computes an intermediate velocity $\textbf{u}^*$ by time stepping and ignoring the pressure term
\begin{align}
    \frac{\textbf{u}^* - \textbf{u}^n}{\Delta t} &= -\textbf{u}^n \cdot \nabla \textbf{u}^n + \nu \nabla^2 \textbf{u}^n
\end{align}
and then ``projecting" the intermediate velocity to the next time step via
\begin{align}
    \textbf{u}^{n+1} = \textbf{u}^* - \frac{\Delta t}{\rho} \nabla P^{n+1}
\end{align}
In order to solve for the right-hand side of the second step, the pressure field at the $n+1$ time step must be known. This is found by solving the following Poisson's equation arising from taking the divergence of the second step and using the conservation of mass to eliminate the $\nabla \cdot \textbf{u}^{n+1}$ term:
\begin{align}
    \nabla^2 P^{n+1} &= \frac{\rho}{\Delta t} \nabla \cdot \textbf{u}^*
\end{align}

This projection method requires an elliptic solve at every time step to solve for the pressure field. In applications like this where an elliptic solve is required multiple times, direct methods where one can pre-compute the solution operator and then apply it each time step greatly speeds up the computation.

\subsection{The Shallow Water Equations and the Serre-Green-Naghdi Model}

\begin{figure}
    \centering
    \includegraphics[width=\columnwidth]{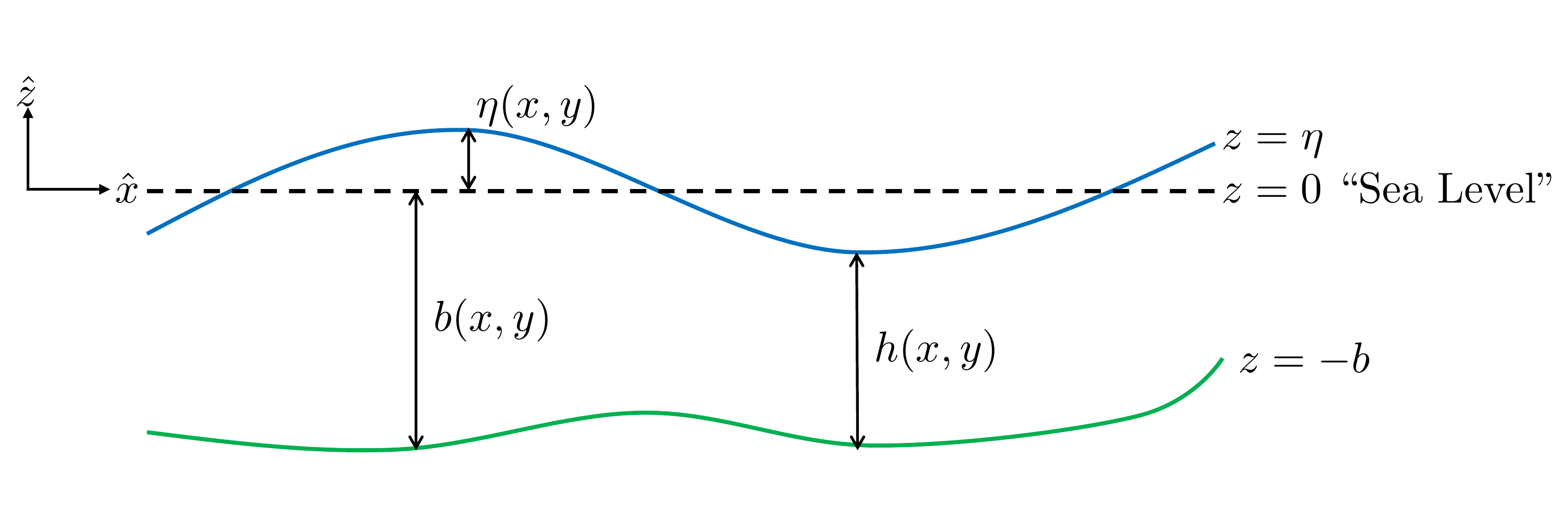}
    \caption{Shallow Water Equation Reference}
    \label{fig:swe}
\end{figure}

The shallow water equations (SWE) form a model for free surface flow derived from potential theory. The key assumption in the SWEs is that the horizontal scale is much larger than the vertical scale. Because a ``shallow limit" is assumed, the velocity field in the vertical direction is a depth-averaged velocity. These types of equations are used to model tsunami waves and debris flow, among other things.

Extensions to this model done by Bonneton et al. in \cite{bonneton2011splitting}, \cite{lannes2009derivation}, and \cite{lannes2015new} introduce a dispersive term correction in the form of a source term, leading to
\begin{align}
    \frac{\partial \textbf{q}}{\partial t} + \nabla \cdot \textbf{F}(\textbf{q}) = \textbf{b} + \textbf{s}
\end{align}
where
\begin{align}
    \textbf{q} &=
    \begin{bmatrix}
        h \\
        h u \\
        h v \\
    \end{bmatrix} \\
    \textbf{F}(\textbf{q}) &=
    \begin{bmatrix}
        h u & h v \\
        h u^2 + \frac{1}{2} gh^2 & h u v \\
        h u v & h v^2 + \frac{1}{2} gh^2 \\
    \end{bmatrix} \\
    \textbf{b} &=
    \begin{bmatrix}
        0 \\
        -g h \frac{\partial b}{\partial x} \\
        -g h \frac{\partial b}{\partial y} \\
    \end{bmatrix} \\
    \textbf{s} &=
    \begin{bmatrix}
        0 \\
        \frac{\alpha - 1}{\alpha} g h \frac{\partial \eta}{\partial x} + d_x \\
        \frac{\alpha - 1}{\alpha} g h \frac{\partial \eta}{\partial y} + d_y \\
    \end{bmatrix}.
\end{align}
The water column height is denoted by $h(x,y)$, with surface velocity $(u(x,y), v(x,y))$, $b(x,y)$ is a bathymetry term (elevation), $\eta(x,y)$ is the free surface height and dispersive term correction $(d_x, d_y)$. Figure \ref{fig:swe} shows the typical setup for this kind of problem.

In order to solve for the dispersive source term, one must solve the following elliptic equation
\begin{align}
    \big( \textbf{I} + \alpha \textbf{T}_{diag}^b \big) \textbf{d} = \textbf{f}(h, \textbf{u}, b, \eta)
\end{align}
where the details of the right-hand side function can be found in \cite{lannes2015new}. The operator $\textbf{T}_{diag}^b$ is an elliptic-like operator that does not change with time. So, for a static mesh, the dispersive operator can be factorized via a direct method and applied at each time step. This reduces the need to solve the elliptic PDE every time step to just an application of the solution operator every time step. If the mesh is adapted, the solution operator would need to be reformed. However, direct methods like the HPS method can allow for a local adjustment to the solution operator, further accelerating the solution.

\section{Open Research Questions}
\label{sec:open}

There remain topics for potential research in the area of fast solvers for elliptic PDEs. In this paper, we looked at classic approaches to approximating solutions of PDEs including finite difference, finite volume, finite element, and spectral methods. Each of these approaches have been analyzed and expanded, with high order methods and fast and parallel algorithms being the primary focus of recent advancements. Although other discretization methods such as collocation methods and meshless methods remain an ongoing research topic, we will look at some open research topics in the area of solution methods, as it is an area with more potential questions and applications. We have already addressed classic solution methods, which we categorized as iterative, direct, or hierarchical. Iterative methods such as splitting or Krylov methods and direct methods such as matrix factorizations are well established in the literature, with textbooks and courses addressing these topics. Indeed, these topics form the foundation for numerical analysis for PDEs. The methods we classified as hierarchical methods (i.e., the multigrid method, nested dissection, and the HPS method) are more open topics. Multigrid methods remain a popular solution technique for a variety of discretization methods, as well as on adaptive meshes (\cite{babich2010adaptive,thompson1989adaptive}). The HPS method has also been introduced and is being improved upon, as in (\cite{fortunato2020ultraspherical,geldermans2019adaptive,gillman2014direct,martinsson2015hierarchical}). However, open questions exist with regard to parallel implementations and further applications of these methods.

Applying the multigrid or HPS method to a particular problem opens up a wide range of new ideas. For example, in the area of computational fluid dynamics (CFD), the multigrid method has been used to help solve the Navier-Stokes equations (\cite{babich2010adaptive}). Wherever the multigrid method has been applied, the HPS method can also be applied for some competitive results. It would be of interest to see how the HPS method and multigrid method compare when applied to the same problem with the same discretization scheme. For example, the multigrid method has been applied to the Serre-Green-Naghdi model with a quadtree discretization (\cite{popinet2015quadtree}). The HPS method works well on quadtrees and can also be applied here, with the advantage of a direct method for faster solutions.

Both the multigrid and HPS method have potential for high scalability. The multigrid method has been implemented in parallel in a variety of papers (\cite{bergen2006massively}). The main approach is to use a domain decomposition approach to divide the domain among processors. The HPS method, however, is fairly new and has limited study in parallelization. Beams et al. in \cite{beams2020parallel} describe a shared-memory implementation of the HPS method. However, they do not address distributed memory paradigms or implementations on a GPU. Additionally, while their results show promising scaling results, they use a relatively small number of cores (up to 28 physical cores). It would be of interest to see how the HPS method scales with communication cost and on much larger machines using a distributed memory paradigm or heterogeneous architecture. The motivation for highly scalable algorithms comes from the advances in high performance computing. Modern supercomputers have millions of cores, and their speed comes from the distribution of work. State of the art supercomputers need to use advanced algorithms that scale well enough to take advantage of the compute power. This is the mission of the United States' Department of Energy's (DoE) Exascale Computing Project (ECP); to combine exascale-capable machines with highly scalable algorithms for ``codesign, modeling and simulation, data analytics, machine learning, and artificial intelligence" (\cite{doe2021exascale}).

\section{Conclusion}
\label{sec:conclusion}

The discretization of elliptic partial differential equations leads to linear systems that can be solved with efficient solvers, taking advantage of the structure of the problem. Discretization methods include the finite difference method, finite volume method, finite element method, spectral methods, and more. These discretizations lead to linear systems of the standard form $\textbf{A} \textbf{u} = \textbf{f}$, where $\textbf{A}$ is sparse and structured. Efficient methods such as splitting and Krylov methods, matrix factorizations, and hierarchical domain decomposition methods can take advantage of the structure of the linear system. These solution methods can also be expanded upon in order to achieve near linear performance and are well suited for high-performance applications.

In addition, we presented various areas of study that could greatly benefit from an efficient, fast, and scalable solver for elliptic PDEs. This includes many applications in computational fluid dynamics like the incompressible Navier-Stokes and the shallow water wave equations. Using projection methods for the Navier-Stokes equations and the Serre-Green-Naghdi dispersive model leads to elliptic terms that must be solved at each time step. Fast, direct methods like the ones currently being studied and implemented can accelerate these areas as well as any area that solves an elliptic partial differential equation.

% =================================================================================================
% bibliography
% =================================================================================================
\bibliographystyle{packages/siamplain}
\bibliography{bib/comp_exam}

% =================================================================================================
% End Document
% =================================================================================================
\end{document}